\documentclass[structabstract]{aa}  
\usepackage{graphicx}
\begin{document}
\title{The growth of supermassive black holes fed by accretion disks}

%\subtitle{}

\author{M.A. Montesinos Armijo \and J.A. de Freitas Pacheco}
\authorrunning{M. Montesinos  \and J.A. de Freitas Pacheco}

\institute{Universit\'e de Nice-Sophia Antipolis\\
Observatoire de la C\^ote d'Azur, Laboratoire Cassiop\'ee, UMR 6202\\
BP 4229, 06304 - Nice Cedex 4 - France\\
\email{matias.montesinos@oca.eu;pacheco@oca.eu}}
\date{\today}

% \abstract{}{}{}{}{} 
% 5 {} token are mandatory
 
  \abstract
  {Supermassive black holes are probably present in the centre of the majority of the galaxies. There is a
consensus that these exotic objects are formed by the growth of seeds either by accreting mass from a 
circumnuclear disk and/or by coalescences during merger episodes.}     
  {The mass fraction of the disk captured by the central object and the related timescale are still open questions,
as well as how these quantities depend on parameters like the initial mass of the disk or the seed or on the
angular momentum transport mechanism. This paper is addressed to these particular aspects of the accretion disk
evolution and of the growth of seeds.}
 {The time-dependent hydrodynamic equations were solved numerically for an axi-symmetric disk in which the gravitational
potential includes contributions both from the central object and from the disk itself. The numerical code is based
on a Eulerian formalism, using a finite difference method of second-order, according to the Van Leer upwind algorithm
on a staggered mesh.}  
  {The present simulations indicate that seeds capture about a half of the initial disk mass, a result weakly dependent
on model parameters. The timescales required for accreting 50\% of the disk mass are in the range 130-540 Myr, depending
on the adopted parameters. These timescales permit to explain the presence of bright quasars at $z \sim 6.5$. 
Moreover, at the end of the disk
evolution, a ``torus-like" geometry develops, offering a natural explanation for the presence of these structures
in the central regions of AGNs, representing an additional support to the unified model.}
   

   \maketitle
%
%________________________________________________________________

\section{Introduction}

At the present time, the general belief is that supermassive black holes (SMBHs) located at the center
of galaxies have been formed by the growth of primordial ``seeds either by matter accretion or
coalescences during merger episodes. This picture is consistent with the fact that
the present black hole (BH) mass density compares quite well with the mass density derived from the
bolometric luminosity function of quasars, under the assumption that the accretion process itself is
the source of the radiated energy (Soltan 1982; Small \& Blandford 1992; Hopkins, Richards 
\& Hernquist 2007). The accreted matter is essentially baryonic in origin
with an eventual contribution of the exotic dark component not exceeding 10\% of the total accreted 
matter (Peirani \& de Freitas Pacheco 2008). 

Nevertheless, the formation of massive BHs by the direct collapse of central regions of proto-galaxies
is an alternative possibility considered by different authors. Shapiro (2004) considered the
core collapse of relativistic star clusters formed in starbursts, which may have occurred in the early 
evolution of galaxies. The collapse of a collisionless system, after having experienced a phase of strong oscillations
followed by violent relaxation, may undergo a slow gravothermal instability, if the phase space 
incompressibility condition is violated (Levin, Pakter \& Rizzato 2008), leading eventually to the
formation of a BH. Despite the non inclusion of the star formation process and its 
associated feedback, Wise, Turk \& Abel (2009) found from hydrodynamical simulations of $\sim 10^8~M_{\odot}$
proto-galaxies, that massive central objects with masses around $10^5~M_{\odot}$ could formed inside regions of one
parsec radius. The most favourable physical conditions for the gas in these proto-halos to form 
those massive central objects are when the virial temperature is about 15000 K and the 
circular velocity is around 20 km/s (Reagan \& Haehnelt 2009).

If the aforementioned investigations open the possibility of forming directly massive BH via the gravitational
collapse, other studies lead to the formation of intermediate mass BH, which could be the required seeds.
Black holes in the mass range ($10^3-10^4 M_{\odot}$) could be formed
in the collapse of primordial gas clouds (Haehnelt \& Rees 1993; Eisenstein \& Loeb 1995; 
Koushiappas, Bullock \& Dekel 2004). Another appealing possibility, which will be considered
in the present investigation, is the formation of BH seeds with masses in the range
$10^2-10^4~M_{\odot}$, resulting from the evolution of primordial massive stars (Bromm, Coppi \& Larson 1999;
Abel, Bryan \& Norman 2000; Heger et al. 2003; Yoshida et al. 2006). These stars would 
have been formed at $z \sim 15-20$
in the high density peaks of the primordial fluctuation spectrum (Gao et al. 2007) and their high masses
would be the consequence of very inefficient gas cooling at zero metallicity. Besides forming BH
seeds, these massive stars could also have contributed to the reionization of the universe.

Different investigations based on cosmological simulations have been performed in these past years,
aiming to understand not only the growth process of seeds, but also the consequences of the feedback
resulting from the accretion process itself on the environment as well as the observed correlations between
the black hole mass with properties of the host galaxy (DiMatteo et al. 2003; Pelupessy, DiMatteo \ Ciardi 2007;
Springel, DiMatteo \& Hernquist 2005; Filloux et al. 2009).
In all these simulations, the physical description of the accretion process requires
a substantial simplification, since it occurs on unresolved spatial scales. If the BH is at  rest with respect
to the gas, the inflow is probably spherical, almost adiabatic and the accretion rate is estimated 
by the Hoyle-Lyttleton-Bondi (HLB) formula 
(Hoyle \& Lyttleton 1939; Bondi \& Hoyle 1944), whose rate depends on parameters like the gas density and 
the sound velocity evaluated far away from the BH, on scales supposed to be resolved by simulations.
The assumption that BHs could be at rest and that the accretion process is spherically symmetric
is probably unrealistic but frequently used in cosmological simulations due to its simplicity. However,
the radiation emitted during the inflow, in particular near the BH horizon, affects the surrounding gas, reducing 
considerably the accretion rate and rendering inefficient the growth of seeds by such a mechanism
(Milosavljevic et al. 2009). The situation is rather different
if the BH is moving with respect to the gas. After passing the BH, a conically shaped
shock is produced in the flow, in which the gas loses the momentum component perpendicular to the shock
front. After compression in the shock, gas particles within a certain impact parameter will fall into
the black hole. Particles having angular momentum exceeding $2r_gc$, with $r_g$ being the gravitational
radius, will form a disk and only after viscous stresses have transported away the excess of
angular momentum will the gas cross the BH horizon. Clearly, all these processes are not
adequately described in any of the aforementioned simulations and one may wonder if the accretion or the
resulting luminosity are estimated satisfactorily when the Hoyle-Lyttleton-Bondi approach is
adopted. 

Numerical simulations indicate that after the merger of two galaxies, a considerable amount of gas is settled
into the central region of the resulting object. The gas loses angular momentum in a timescale comparable
to the dynamical timescale, forming a circumnuclear self-gravitating disk (Mihos \& Hernquist 1996;
Barnes 2002), having masses in the range $10^6-10^9~M_{\odot}$ and dimensions of about $100-500~pc$. Such a
disk will probably be able to feed the central BH, increasing its mass. In fact, this scenario was considered
by Filloux et al. (2009) in their simulations. These authors adopted an accretion rate derived from a steady disk model,
in which the angular momentum transport is provided by turbulent viscosity. Since in reality the disk is 
expected to be non-steady, they assumed that during a time step the properties of the disk do not change 
appreciable but they are updated continuously at each new time step. Despite this simple description of the accretion process,
they were able to reproduce quite satisfactory the main properties of BHs and their host galaxies. More
recently, Power, Nayakshin \& King (2010) introduced a new method to model the physics of an accreting
disk around a BH, which has some common points with  the approach by Filloux et al. (2009). Based on a series
of numerical experiments, they have shown that the disk accretion mode is more physically consistent than
the simple use of the HLB formula and more efficient to feed the seed, in agreement with
the conclusions by Filloux et al. (2009). 
In these last two approaches, the accretion disk is represented in simulations by 
a ``single" (eventually two in  rare cases) SPH particle and the adopted prescription for
the accretion rate defines the mass fraction transferred to the central BH. In the real world, the
structure of the disk changes continuously for different reasons: mass is transferred either to
central BH or to the outskirts of the disk, stars form consuming part of gas available 
to feed the BH and infall of matter is a source of fresh gas replenishing the disk. 

The understanding of the aforementioned aspects require a close inspection
of the accretion process, in order to improve the modelization adopted in cosmological simulations. 
In the present paper we report results on the numerical solution of the hydrodynamic equations 
describing the evolution of a circumnuclear disk around a black hole, aiming to answer some specific questions 
that could be useful to improve the modelling of such a process in cosmological simulations. Moreover, SLOAN
quasars associated to very massive BHs (Fan et al. 2001; Willot, McLure \& Jarvis 2003) are already 
observed at redshifts $z \sim 6$, when the universe was only 0.96 Gyr old. These observation imply that 
the growth process of seeds should occur in shorter 
timescales, in regions where the accretion mechanism must be very efficient. The present investigation
will be mainly focused in the following questions: 1) what is the fraction of the disk mass 
accreted by the central BH? 2) Does such a fraction depend on the initial mass either of the BH seed 
or the disk? 3) What are the typical accretion timescales and how they depend on the disk parameters?
This paper is organized as follows: in Section 2 the accretion disk model is discussed
and the main equations are introduced; in Section 3 the numerical methods are presented; in Section 4
the main results are given and, finally, in Section 5 the conclusions are summarized.

\section{The accretion disk model}

In this work, we will study the evolution of nuclear disks formed possibly after a merger
event or by gas infall, having masses in the range $10^7 - 10^8~M_{\odot}$ and with typical dimensions of the
order of 50 pc. These disks are initially self-gravitating and, as the central BH
grows, it dominates the dynamics of the inner regions while the outer parts
are still dominated by the disk self gravitation. As we shall see later, the external disk
is constituted by neutral gas having temperatures of about $100-2000~K$, conditions favourable to
form stars. However, in the present work, the gas conversion into stars through the disk will not be 
considered, although a new version of our code, presently under development, will 
offer such a possibility.  

\subsection{Angular momentum transport}

A major problem in the current understanding of accretion disks is the absence of
a physical theory able to describe the viscosity of the gas in the presence of
turbulent flows or in the presence of a magnetic field. The angular momentum transport
in accretion disks is generally described by the formalism introduced almost forty
years ago by Shakura \& Sunyaev (1973), in which the viscosity due to the subsonic
turbulence is parametrized by the relation
\begin{equation}
\eta =\alpha Hc_s
\label{alpha}
\end{equation}
where $\alpha \leq 1$ is a dimensionless coefficient, $H$ is the vertical scale of the disk, supposed to
be of the same order as the typical (isotropic) turbulence scale $\ell_t$ and $c_s$ is the sound velocity.
Accretion disk models based on the ``$\alpha$-viscosity" approach are able to explain successfully the observed
properties of dwarf-novae and X-ray binaries, but disks based on such a formulation
are thermally unstable as demonstrated long time ago by Piran (1978). 

The possibility that turbulence could be generated by local gravitational instability in geometrically
thin disks was considered by Paczynski (1978). Duschl and Britsch (2006) revisited this idea 
by analysing the gravitational instability of self-gravitating disks, concluding that such an instability 
could be able to develop turbulence in the flow and, consequently, to generate viscosity without requiring 
the contribution of magneto-hydrodynamic turbulence. 
Recent simulations of the gas inflow in the central regions of galaxies, induced by the
gravitational potential either of the stellar nucleus or the SMBH,
reveal the appearance of highly supersonic turbulence, with velocities of the order
of the virial value (Regan \& Haehenelt 2009; Levine et al. 2008; Wise, Turk \& Abel 2008). 
Amazingly, no fragmentation is observed in such a gas despite of being isothermal and gravitationally unstable.
In fact, Begelman \& Shlosman (2009) argued that an efficient angular momentum
transfer suppresses fragmentation, favouring the gas inflow. On the contrary, if the angular momentum transfer is
inefficient, the turbulence decays and triggers global instabilities which regenerates
a turbulent flow. Moreover, according to their analysis, fragmentation is suppressed whenever the gas
temperature remains below the virial value.  As we shall see
later, during the early evolutionary phases of the disk, the accretion rate is quite important. 
Consequently, the heat generated by dissipation of turbulence increases the radiation pressure which
inflates the inner regions, producing a ``slim" disk.
We would expect that a flow self-regulated by the aforementioned mechanism could be stable
against fragmentation. If the flow is self-regulated, it must be characterized by a critical
Reynolds number ${\cal R}$, determined by the viscosity below which the flow becomes unstable.
In fact, de Freitas Pacheco \& Steiner (1976) suggested that instead of the ``$\alpha$" parametrization,
the effective (turbulent) viscosity $\eta$ could be expressed in terms of 
such a critical Reynolds number by
\begin{equation}
\eta = \frac{2\pi rV_{\phi}}{\cal R}
\label{reynolds}
\end{equation}
where $r$ is the radial distance to the center of the disk (or to the central BH) and $V_{\phi}=\Omega r$ 
is the azimuthal velocity of the gas. It is worth mentioning that this formulation, which will be adopted here,
is essentially the ``$\beta$-viscosity" model discussed by Duschl, Strittmatter \& Biermann (1998) (see also
Richard \& Zahn 1999). An additional aspect justifying the adoption of this formulation is that 
accretion disks modeled by such a viscosity prescription are thermally stable according to the
analysis by Piran (1978).  However, in the present
work the disk stability was not investigated and we cannot exclude the possibility that fragmentation occurs.
This aspect will be examined in a future paper.

It can be verified trivially that in such a formulation the Mach number
associated to the mean turbulent motions is $M_t \simeq V_{\phi}/(c_s\sqrt{\cal R})$. This implies
that turbulence is slightly supersonic in most regions of the disk but it can be supersonic at the inner
regions, depending on the critical Reynolds number.

\subsection{The dynamical equations}

The hydrodynamic equations are written in cylindrical coordinates ($r, \phi, z$) and, in order to simplify the solution
of the Poisson equation, variables are integrated along the z-axis but the scale of height is calculated in
a consistent way as we shall see below. The equation of the radial motion, neglecting gas pressure gradients, is
\begin{equation}
\frac{\partial V_r}{\partial t}+V_r\frac{\partial V_r}{\partial r}-\frac{V^2_{\phi}}{r}-
\frac{\partial\psi}{\partial r}=0
\label{radial}
\end{equation}
where $V_r$ is the radial velocity, $V_{\phi}$ is the azimuthal velocity and $\psi$ is the gravitational potential,
including the contribution of the central BH and of the disk itself. The central BH is probably rotating but in the
present investigation we discard this possibility and we use the approximate potential of Paczy\'nski-Wiita (Paczy\'nski
\& Wiita 1980) to model the gravitational field of the BH. The contribution of the disk itself is obtained from the
expression for the internal potential of an ellipsoid in which we have performed 
the limit $z \rightarrow 0$. Under these conditions
the total gravitational force is
\begin{equation}
\frac{\partial\psi}{\partial r}=-\frac{GM_{BH}(t)}{(r-r_g)^2}-\frac{G}{r}\int^r_0\frac{\Sigma(a,t)ada}{\sqrt{r^2-a^2}}
\label{potencial}
\end{equation}
where $r_g = 2GM_{BH}/c^2$ is the gravitational (or the event horizon) radius and $\Sigma$ is the 
columnar mass density defined by 
\begin{equation}
\Sigma(r,t) = \int^{\infty}_{-\infty}\rho(r,z,t)dz
\label{density}
\end{equation}
The mass of the BH varies according to the equation
\begin{equation}
\frac{dM_{BH}(t)}{dt}=2\pi r_{lso}(t)\Sigma(r_{lso},t)V_r(r_{lso},t)
\end{equation}
where the quantities on the right side are evaluated at the radius of the last stable circular orbit $r_{lso}$. Notice
that as the BH grows, the radius of the horizon and of the last stable circular orbit increase 
proportionally to the BH mass.

The continuity equation is given by
\begin{equation}
\frac{\partial\Sigma}{\partial t}+\frac{1}{r}\frac{\partial(r\Sigma V_r)}{\partial r}=0
\label{continuity}
\end{equation}
and the equation for the azimuthal motion, including the viscous forces responsible for
the angular momentum transport is
\begin{equation}
\frac{\partial V_{\phi}}{\partial t}+V_r\frac{\partial V_{\phi}}{\partial r}+\frac{V_rV_{\phi}}{r}+
\frac{1}{r\Sigma}\frac{\partial(rT_{r\phi})}{\partial r}=0
\label{tangencial}
\end{equation}
where the considered component of the stress tensor (integrated along the z-axis) is
\begin{equation}
T_{r\phi}=\eta\Sigma r\frac{\partial\Omega}{\partial r}
\end{equation}
and, using eq.~\ref{reynolds}, the equation above can be recast as
\begin{equation}
T_{r\phi}=\left(\frac{2\pi}{{\cal R}}\right)\Sigma r^3\Omega\frac{\partial\Omega}{\partial r}
\end{equation}

\subsection{The scale of height}

Assuming that the disk is in hydrostatic equilibrium along the z-axis, we can write
\begin{equation}
\frac{d(P_g+P_t+P_r)}{dz}=-\rho g_z 
\label{hydrostatic}
\end{equation}
where $P_g = \rho c^2_s$, $P_t = \rho <v_t^2>$ and $P_r = aT^4/3$ are respectively the gas, the turbulent and the
radiation pressure. The vertical component of the gravitational acceleration has 
two contributions: one from the disk self-gravity and another from the tidal force due 
to the central black hole. Thus,
\begin{equation}
g_z = 2\pi G\Sigma\frac{z}{H}+ \frac{GM_{BH}}{r^3}z
\label{vertical}
\end{equation}
Defining an effective scale of height by the relation $H=\rho/(\mid d\rho/dz\mid)$ and using the equations above, one 
obtains after some algebra
\begin{equation}
H=-\frac{c_s}{\bar Q\Omega_K}\left[(1-\beta)-\sqrt{(1-\beta)^2+\bar Q^2(1+\varepsilon^2)}\right]
\label{height}
\end{equation}
where $\varepsilon^2=<v_t^2>/c_s^2$ and we have introduced  respectively
\begin{equation}
\Omega_K^2=\frac{GM_{BH}}{r^3}
\end{equation}
and
\begin{equation}
\bar Q = \frac{\Omega_Kc_s}{\pi G\Sigma}
\end{equation}
The parameter $\beta$ defined as
\begin{equation}
\beta = \frac{aT^4}{3\pi G\Sigma^2}
\end{equation}
measures the ratio between the radiation pressure and the disk self-gravity. 
Notice that from eq.~\ref{height} some limiting cases can be derived. When the radiation force is unimportant
($\beta << 1$) and self-gravity dominates ($\bar Q << 1$), one obtains
\begin{equation}
H \simeq \frac{c_s^2(1+\varepsilon^2)}{2\pi G\Sigma}
\end{equation}
whereas when $\bar Q >> 1$, namely, the vertical acceleration is dominated by tidal forces we have
\begin{equation}
H \simeq \frac{c_s\sqrt{(1+\varepsilon^2)}}{\Omega_K}
\end{equation}
It is worth mentioning that when the disk becomes optically thin, the contribution due to the radiation 
pressure is negligible and, in this case the scale of height is computed from eq.~\ref{height} with 
the condition $\beta$=0. 

\subsection{The temperature of the disk}

We assume that all energy is generated locally by viscous dissipation and radiated essentially along the z-axis
although a small fraction could be advected inwards. Under these conditions, the local energy balance is given by
\begin{equation}
Q_{dis}=Q_{rad}+Q_{adv}
\label{balance}
\end{equation}
The energy rate per unit of area dissipated by viscous forces is  
\begin{equation}
Q_{dis}=\frac{2\pi}{{\cal R}}\Sigma\Omega^3r^2\left(\frac{d\lg\Omega}{d\lg r}\right)^2
\label{viscous}
\end{equation}
while the advected energy flux is (Abramowicz et al. 1986)
\begin{equation}
Q_{adv}=f(r)\left(\frac{H}{r}\right)^2Q_{dis}
\end{equation}
where $f(r)$ depends on the angular momentum distribution throughout the disk but, in general, it is of the order of the
unity (Abramowicz et al 1995), except near the star-disk transition layer, where it could 
attain quite large values. In our case, since
the central object is a black hole, such a layer doesn't exist and the $f(r)$ factor depends
essentially on the difference between the angular momentum at the considered position and that at the 
last stable orbit. Here we assume simply that $f(r)=1$ everywhere in the disk.

In the regions where the disk is partially or completely ionized, the opacity is due to the Thomson scattering,
bound-free and free-free transitions. In this case, using the solution of the transfer equation within the 
Eddington approximation derived by de Freitas Pacheco, Steiner \& Daminelli (1977), the local emergent flux is
\begin{equation}
\Phi_{\nu}(T_e(r))\simeq \frac{2\pi B_{\nu}(T_e)}{\left[1+\frac{\sqrt{3}}{2\lambda_{\nu}}coth t_{0,\nu}\right]}
\label{transfer1}
\end{equation} 
where $t_{0,\nu}=\sqrt{3\tau_f\left(\tau_f+\tau_s\right)}$ is the effective 
optical depth, with $\tau_f$ and $\tau_s$ being
respectively the absorption and scattering optical depths. The other parameter is defined by the relation
$\lambda^2_{\nu}=\tau_f/(\tau_f+\tau_s)$. $T_e$ is the effective temperature. Thus,
\begin{equation}
Q_{rad}= 2\int^{\infty}_0\Phi_{\nu}(T_e(r))d\nu
\end{equation}
where the factor $2$ takes into account both disk surfaces.

Another factor affecting the evaluation of the disk temperature is the photon trapping. At very high optical depths,
the photon diffusion timescale along the z-axis $t_d = 3(H/c)t_0$  can be larger than the accretion 
timescale $t_{ac} = r/V_r$. Here $t_0$ is the frequency averaged effective optical depth as defined above.
When this occurs, photons are not able to reach the surface, being convected inwards and 
swallowed by the central BH. In this case, besides the advection correction, the dissipated energy should be 
reduced by an extra factor equal to the ratio $t_{ac}/t_d$.

The local effective temperature is derived from the numerical solution of the above equations.
In the outer regions, the disk is neutral and optically thin. In this case, the temperature was
estimated simply from the balance between the heating and the cooling rates. The cooling is due essentially
to molecular hydrogen in the case of a disk constituted essentially by primordial gas. When trace elements
are present, an additional cooling mechanism should be considered, which is due to the excitation
of fine-structure levels of $C^+$, $Si^+$, $O^0$ and $Fe^+$. The chemical reactions leading to the formation
of the $H_2$ molecule requires a residual electron density, which we will assume to be provided by
the ionizing radiation background. Under this conditions, we have adopted the
cooling functions for molecular hydrogen given by Galli \& Palla (1998) and those for atomic infrared lines 
by de Freitas Pacheco (1969).

\section{The numerical method}

In order to solve numerically the hydrodynamic equations describing the structure and the evolution of the disk,
we have modified the public code FARGO (Masset 2000), originally developed for studies of the interaction between
planets and circumstellar disks. In fact, the structure of this code is quite similar to ZEUS (Stone \& Norman 1992).
The code is based on an Eulerian formalism, using a finite difference method of second-order according to
the Van Leer upwind algorithm on a staggered-mesh (Van Leer 1977), which means that scalar quantities like
surface density, scale of height, etc., are defined at the center of the cell whereas vector quantities like
velocity, fluxes in general, etc., are defined at the interface between cells. In our particular case, the
grid implementation was done in a logarithmic scale, since the radial scale may vary between nine up to eleven
orders of magnitude in order to cover all the extension of the disk from the last stable orbit until the
external radius. Such a logarithmic grid was covered by 1024 ring sectors.

The time-step is controlled by the Courant-Friedrich-Levy (CFL) condition, which states that the information cannot
sweep a distance larger than the size of a cell $(R_i-R_{i-1})$ over one time-step. This condition introduces five
constraints, corresponding respectively to five time-step 
limits $\delta t_1$, $\delta t_2$, $\delta t_3$, $\delta t_4$ and
$\delta t_5$. The first one is defined by
\begin{equation}
\delta t_1 = min(\delta r, r\delta\phi)/c_s
\end{equation}
where $\delta r$ and $r\delta\phi$ correspond to the radial and azimuthal mesh size. This condition implies that
no wave propagating with velocity $c_s$ should cross the cell in one time-step. The following two conditions
constrain the motion of a test particle that should not travel a distance greater than $\delta r$ in the radial
direction and $r\delta\phi$ in the azimuthal direction. These conditions are expressed as
\begin{equation}
\delta t_2 = \delta r/V_r
\end{equation}
and
\begin{equation}
\delta t_3 = r\delta\phi/V_{\phi}
\end{equation}  
The fourth condition is related to the artificial viscosity, introduced to avoid large discontinuities due to
the eventual presence of shocks (Richtmyer \& Morton 1957) and is given by
\begin{equation}
\delta t_4 = min\left(\frac{\delta r}{4C^2_v\delta V_r}, \frac{r\delta\phi}{4C^2_v\delta V_{\phi}}\right)
\end{equation}
where $\delta V_r \equiv \partial_rV_r\delta r$ and $\delta V_{\phi}\equiv \partial_{\phi}V_{\phi}\delta\phi$.
Notice that for axial symmetry this last term is zero.
$C_v$ is the Von Neumann-Richtmyer viscosity constant, usually taken in the range (1, 2). Here,
we have assumed $C_v=1.41$ in all runs. The last condition assures that two neighboring rings are not disconnected
after a time-step and it can be expressed as
\begin{equation}
\delta t_5 = \frac{2\pi}{\left[\Omega(r)-\Omega(r+\delta r)\right]}
\end{equation}
Finally, the hydrodynamic time-step is computed from the relation
\begin{equation}
\Delta t = C_{CFL}\times min\left[\frac{1}{\sqrt{\delta t_1^{-2}+\delta t_2^{-2}+\delta t_3^{-2}+
\delta t_4^{-2}}}, \delta t_5\right]
\end{equation}   
where $C_{CFL}$ is the so-called CFL number, usually taken in the range (0, 1) and assumed here to be equal 0.5
in all runs.

In order to understand the different steps of the algorithm, we recall that the hydrodynamic
equations can be written in a general form, i.e.,
\begin{equation}
\partial_tX + V_i\partial_iX = S
\end{equation}
where $X$ is a generic variable (density, mass flux or velocity) and $S$ is a generic source term. Our 
algorithm proceeds in the following way. After evaluation of the gravitational potential, the velocity field is 
updated in each cell, using the source terms and a first order integrator such as: X(t+dt)=X(t)+Sdt. Then, the
considered variable is advected from each cell to its neighbour with a second order upwind interpolation. The
quantities $\Sigma$, $\Sigma V_r$ and $r\Sigma V_{\phi}$ are updated by taking into account their associated
fluxes.

The boundary conditions are simple equations which assign values to the dependent variables in the ghost zones
(inner and outer rings of the grid) corresponding to values in the adjacent ring. The fundamental rule applied
in the ghost zones is that matter are not allowed to enter in the grid zone. Only outflows are permitted.
Thus, in the inner ghost zone (ring zero), the density is that of the 
neighbouring ring (ring one). The radial velocity
of the ring one is set equal to that of ring 2 if the later is negative (outflow) and equal to zero otherwise.
A symmetric algorithm is used for the outer ghost zone.

For the initial configuration, we assume that the columnar density has a profile given by
\begin{equation}
\Sigma(r,t=0)=\Sigma_0\frac{r_0}{r}
\end{equation}
corresponding to an initial total disk mass $M_d=2\pi\Sigma_0r_0R_d$, where $R_d$ is the external radius of the disk. 
Supposing that the disk is initially in equilibrium, the considered density distribution implies that the 
initial profile of the azimuthal velocity is
\begin{equation}
V_{\phi}^2 = \frac{GM_{BH}(0)}{(r-r_g)}+\frac{\pi}{2}G\Sigma_0r_0
\end{equation}
while the initial profile of the radial velocity is simply $V_r(r,t=0)=0$.

We have adopted the parsec as the unit of length, the initial BH mass as the unit of mass and
the unit of time is 
\begin{equation}
\sqrt{\frac{r_0^3}{GM_{BH}(0)}} = 1.494\times 10^6\left(\frac{r_0}{1pc}\right)^{3/2}
\left(\frac{100M_{\odot}}{M_{BH}(0)}\right)^{1/2} yr
\end{equation}
Therefore, with these units, velocities are given in $0.6545(M_{BH}(0))^{1/2}(1pc/r_0)^{1/2}~km/s$.

If the external radius of the disk is fixed at 50 pc, models are characterized by three parameters:
the masses of the seed and of the disk as well as the critical Reynolds number. Table 1 gives
the parameters of the different models investigated. The first column identifies the model, the second
indicates the disk mass, the third gives the seed mass and the fourth gives the critical Reynolds number.
%%%%%%%%%%%%%%%%%%%%%%%%%%%%%%%%%%table 1%%%%%%%%%%%%%%%%%%%%%%
\begin{table}
\caption{Disk model parameters}            
%\label{table:1}     
\centering                         
\begin{tabular}{c c c c}       
\hline\hline                 
Model&Disk mass &Seed mass &Reynolds number\\
     & $(M_{\odot})$& $(M_{\odot})$&   \\   
\hline                        
   1 & $5\times 10^7$ & 100 & 500\\     
   2 & $5\times 10^8$ & 100 & 500\\
   3 & $5\times 10^8$ & 1500& 500\\
   4 & $5\times 10^7$ & 1500& 500\\
   5 & $5\times 10^7$ & 1500& 1500\\
   6 & $5\times 10^7$ & 100& 1500\\ 
\hline                                   
\end{tabular}
\end{table}
%%%%%%%%%%%%%%%%%%%%%%%%%%%%%%%%%%%%%%%%%%%%%%%%%%%%%%%%%%%%%%%%%%%%%%%

All computations were performed at the Centre of Numerical Computation of the Observatoire
de la C\^ote d'Azur (SIGAMM).

\section{Results}

\subsection{The growth of seeds}

The analysis of the numerical solutions obtained for the considered models indicate that after
$\sim 1.2$ Gyr, the disks have conserved only $\sim$ 1-2\% of their original masses. This result
depends weakly on the initial mass either of the disk or of the seed and are valid for
critical Reynolds numbers in the range $500 - 1500$. About 52-54\% of the disk mass
is accreted by the central BH whereas approximately 45\% are ``lost" as a consequence of the expansion of 
the outer regions, due to the redistribution of angular momentum throughout the disk. These values are
probably upper limits since the star formation process, not included in the present approach, may consume a
substantial fraction of the gas.

%%%%%%%%%%%%%%%%%%%%%%%%%%%%%%%%%%%%%%%%%% fig 1 %%%%%%%%%%%%%%%%%%%%%%%%%%%%%%%%%%%%%%%%%%%%%%%%%
\begin{figure}
\centering
\includegraphics[width=8cm]{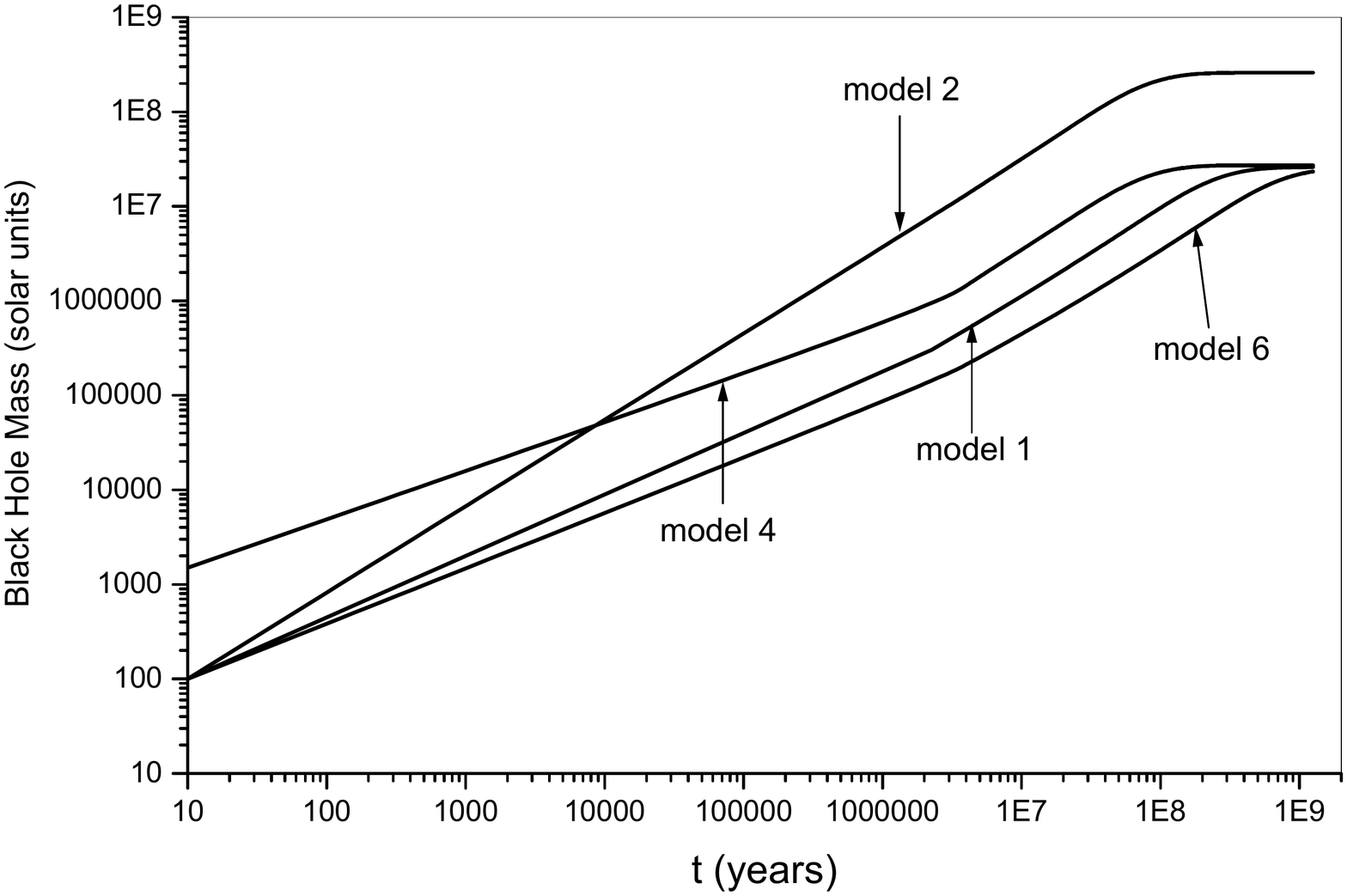}
\caption{Evolution of the black hole masses. Labels indicate the considered
models, whose parameters are given in table 1.}
%\label{FigI}
\end{figure}
%%%%%%%%%%%%%%%%%%%%%%%%%%%%%%%%%%%%%%%%%%%%%%%%%%%%%%%%%%%%%%%%%%%%%%%%%%%%%%%%%%%%%%%%%%%%%%%%%%

Figure 1 shows the evolution of the central black hole mass for models 1, 2, 4 and 6, since models 3 and 5
have a behavior respectively similar to models 2 and 1. A close inspection of these curves permits an
evaluation of the parameter $t_{50}$, which measures the timescale required for the seed to accrete 
50\% of the initial disk mass. Table 2 gives for each model the mass fraction accreted
by the central BH after 1.2 Gyr and the parameter $t_{50}$ in Myr. The mass fraction of the disk beyond
the initial radius (50 pc) or the mass ``lost"  due to the expansion is also given in table 1 
as well as the parameter $t_{40}$, which measures the timescale necessary for the disk to 
lose 40\% of its initial mass due to the redistribution of angular momentum. Notice that for
model 6, which has a seed of 100 $M_{\odot}$ and a critical Reynolds number equal to 1500, the timescale
$t_{50}$ is longer than 1.2 Gyr, time interval in which the evolution of the disks was followed and, consequently,
by that time the BH is still growing but at a very slow rate. 

%%%%%%%%%%%%%%%%%%%%%%%%%%%%%%%%%%table 2%%%%%%%%%%%%%%%%%%%%%%
\begin{table}
\caption{Accretion and mass loss parameters: fraction of the initial
disk mass accreted by the seed after 1.2 Gyr (column 2); timescale in Myr for accreting
50\% of the initial disk mass (column 3); fraction of the initial disk
mass ``lost" by expansion (column 4) and timescale in Myr for ``losing"
40\% of the initial disk mass (last column)}            
%\label{table: 2}     
\centering                         
\begin{tabular}{c c c c c}       
\hline\hline                 
Model&Accreted mass&$t_{50}$&Mass lost&$t_{40}$\\   
\hline                        
1 &0.519&541&0.466&157\\     
2 & 0.518&173&0.466&56\\
3 & 0.518&173&0.466&56\\
4 & 0.542&133&0.442&94\\
5 & 0.519&533&0.466&156\\
6 & 0.468&1426&0.462&385\\ 
\hline                                   
\end{tabular}
\end{table}
%%%%%%%%%%%%%%%%%%%%%%%%%%%%%%%%%%%%%%%%%%%%%%%%%%%%%%%%%%%%%%%%%%%%%%%

Models 1 and 6 differ only by the adopted critical Reynolds number. Since the viscous timescale
$t_{vis} = r^2/\eta \sim {\cal R}/\Omega$ is related with the accretion timescale, one should expect that
$t_{50}$ for model 6 would be about three times higher than for model 1, what is in fact 
verified from the numbers given in table 2. Notice that the timescale $t_{40}$ for model 6 is also
larger than that of model 1 by a similar factor. Models 4 and 5 also differ with respect to
the critical Reynolds number as models 1 and 6. However, the comparison between models 4 and 5
indicates that ratio between the values of $t_{50}$ for these models is slightly higher 
than the value expected simply from
the ratio between the critical Reynolds number defining each model. The reason is due to fact
that these models have seed masses considerably higher than models 1 and 6. A higher seed
mass dominates earlier the dynamics of the inner part of the disk, increasing the accretion rate and 
reducing $t_{50}$. Thus, the rapidity at which the central BH accretes 50\% of the disk mass
depends not only of the critical Reynolds number (certainly the main parameter controlling
the accretion process) but also on the initial seed mass. Models 1 and 2 have the same seed mass and
critical Reynolds number but the disk of the latter is ten times more massive than that of
model 1. More massive the disk is, higher the mass density, which leads to a higher accretion rate.
These effects can be seen in fig. 2, in which the evolution of the accretion rate for the different
models is shown. For all models, in the early evolutionary phase, the accretion rate decreases
slightly as the inner parts of the disk are consumed, except for model 4, in which a small increase
is observed up to $t = 8\times 10^5$ yr, followed by a slowly reduction of the accretion rate. At the 
end of this phase, the accretion
rate decreases abruptly and the growth of the BH ceases as well as its activity.

%%%%%%%%%%%%%%%%%%%%%%%%%%%%%%%%%%%%%%%%%% fig 2 %%%%%%%%%%%%%%%%%%%%%%%%%%%%%%%%%%%%%%%%%%%%%%%%%
\begin{figure}
\centering
\includegraphics[width=8cm]{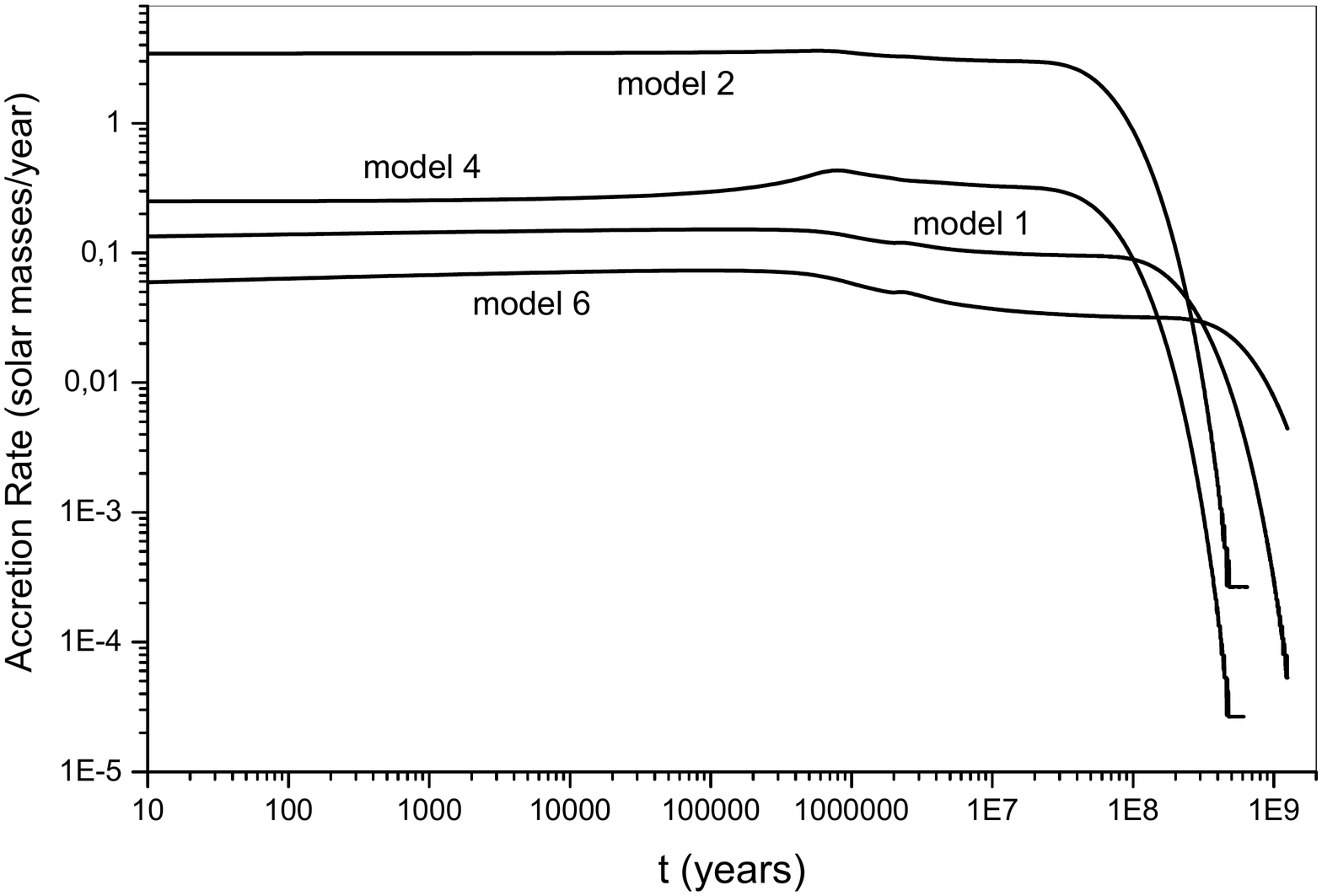}
\caption{Accretion rate evolution for models 1, 2, 4 and 6}
%\label{FigII}
\end{figure}
%%%%%%%%%%%%%%%%%%%%%%%%%%%%%%%%%%%%%%%%%%%%%%%%%%%%%%%%%%%%%%%%%%%%%%%%%%%%%%%%%%%%%%%%%%%%%%%%%%

\subsection{The luminosity evolution}

During the early active phase, the BH can be identified as an AGN or a quasar. Such a phase, whose
timescale is given approximately by $t_{50}$ lasts for 130-540 Myr, according to our models, excepting
for case 6, which has a longer duration due to the smaller accretion rate. These timescales are
consistent with estimates of the duration of the activity based
on the statistics of QSOs and AGNs. For instance, Marconi et al. (2004) estimated that the activity phase
associated to SMBH with masses around $4\times 10^8~M_{\odot}$ is about $450$ Myr while the timescale
activity related to more massive BHs ($\sim 10^9~M_{\odot}$) is considerably shorter, i.e., $\sim 150$ Myr, 
in agreement with our expectations.

Figure 3 shows the evolution of the bolometric luminosity
for some of our disk models, which follows, as expected, the same trend of the 
accretion rate history. Maximum bolometric luminosities
are in the interval $4\times 10^{46}-8\times 10^{44}$ erg/s, well within the observed range. Notice that
a more massive disk ($\sim 4-5\times 10^9~M_{\odot}$) would produce still higher luminosities. The conversion
efficiency of gravitational energy into radiation derived for our models from the ratio $\varepsilon = L/\dot Mc^2$,
indicates that such a quantity does not remain constant during the disk evolution. In the beginning, it is close to the
maximum allowed value ($\varepsilon\sim 1/12$) and then it decreases steadily to values of the order of
$10^{-3}$ after 1.2 Gyr.

In the early evolutionary phases (t $<$ 100 Myr), the disk luminosity exceeds the Eddington value.
This question has already been extensively discussed in the literature, since one may wonder if such a limit 
should be applied to accretion disks (see, for instance, Heinzeller \& Duschl 2007 for a recent discussion on this
subject). In fact, the 
Eddington limit expresses the maximum radiative flux 
able to cross the outer layers of a star without destroying its hydrostatic equilibrium. In general, the gas 
is supposed to be completely ionized and only the scattering of photons by free electrons is taken into account
in the interaction between radiation and matter. Under these conditions, the Eddington luminosity
depends only on the mass of the star. In the case of an accretion disk, the geometry is rather different and
the condition for having equilibrium along the z-axis is local. In the region close to the last stable circular
orbit, tidal forces due to the central black hole balance pressure gradients along the z-axis of the disk, maintaining
the hydrostatic equilibrium. If radiation pressure is supposed not to surpass gravity, then the following local
condition should be satisfied
\begin{equation}
Q_{rad}(r) < \frac{cGM_{BH}}{2r^2}\frac{\Sigma}{t_{ef}}\left(\frac{H}{r}\right)
\end{equation}  
Notice that the radiative flux is fixed by eq.~\ref{balance} and, as the disk is inflated by the radiation pressure,
the energy fraction advected inwards increases, permitting higher accretion rates without rising the radiative flux.
This ``cooling" effect is amplified by the photon trapping mechanism, as already remarked by Begelman (1978) and
Ohsuga et al. (2002), contributing to maintain the hydrostatic equilibrium along the vertical axis.

%%%%%%%%%%%%%%%%%%%%%%%%%%%%%%%%%%%%%%%%%% fig 3 %%%%%%%%%%%%%%%%%%%%%%%%%%%%%%%%%%%%%%%%%%%%%%%%%
\begin{figure}
\centering
\includegraphics[width=8cm]{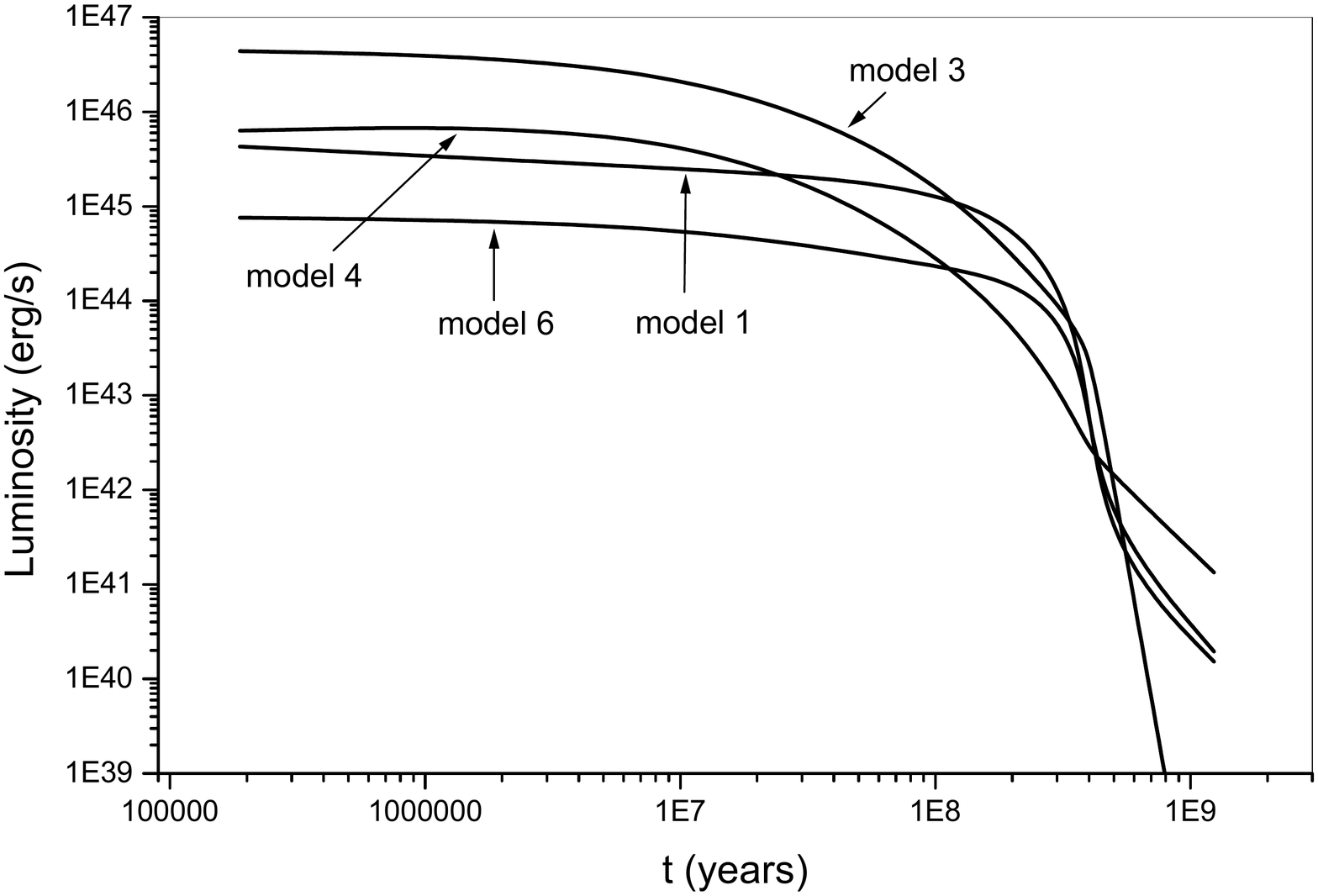}
\caption{The bolometric luminosity evolution for models 1, 3, 4 and 6}
%\label{FigIII}
\end{figure}
%%%%%%%%%%%%%%%%%%%%%%%%%%%%%%%%%%%%%%%%%%%%%%%%%%%%%%%%%%%%%%%%%%%%%%%%%%%%%%%%%%%%%%%%%%%%%%%%%%

\subsection{Physical properties of the disk}

As mentioned previously, the radiation pressure inflates the inner region, producing a slim disk. This
effect is illustrated in figure 4, in which the profile of the aspect ratio $H/r$ for model 4
is shown for three different evolutionary stages. At 9.29 Myr after the beginning of the accretion process,
the radiation pressure is responsible for the increase of the scale of height inside a region  
$r < 5\times 10^{12}$ cm. At this moment, the effective temperature at the inner radius attains a value of about 
$1.8\times 10^6$ K. After 436.5 Myr, the radiation pressure still affects the inner region, since the effective temperature  
is still of the order of $\sim 10^6$ K. Near the end of the evolution, after 1.23 Gyr, when all the disk mass was 
practically consumed, effects of the radiation pressure are no more seen and the scale of height increases
with the distance as expected. The same behaviour is seen for all models having a critical Reynolds number
equal to 500. For models with higher values of ${\cal R}$ (models 5 and 6) this not happens. Higher Reynolds numbers
decrease the radial velocity, increasing the local mass density and decreasing
the parameter $\beta$. Thus, besides the tidal field of the BH, the disk self-gravitation also contributes to maintain
the hydrostatic equilibrium in the inner region, avoiding an important inflation of such a zone as in other
models with lower values of ${\cal R}$.

%%%%%%%%%%%%%%%%%%%%%%%%%%%%%%%%%%%%%%%%%% fig 4 %%%%%%%%%%%%%%%%%%%%%%%%%%%%%%%%%%%%%%%%%%%%%%%%%
\begin{figure}
\centering
\includegraphics[width=8cm]{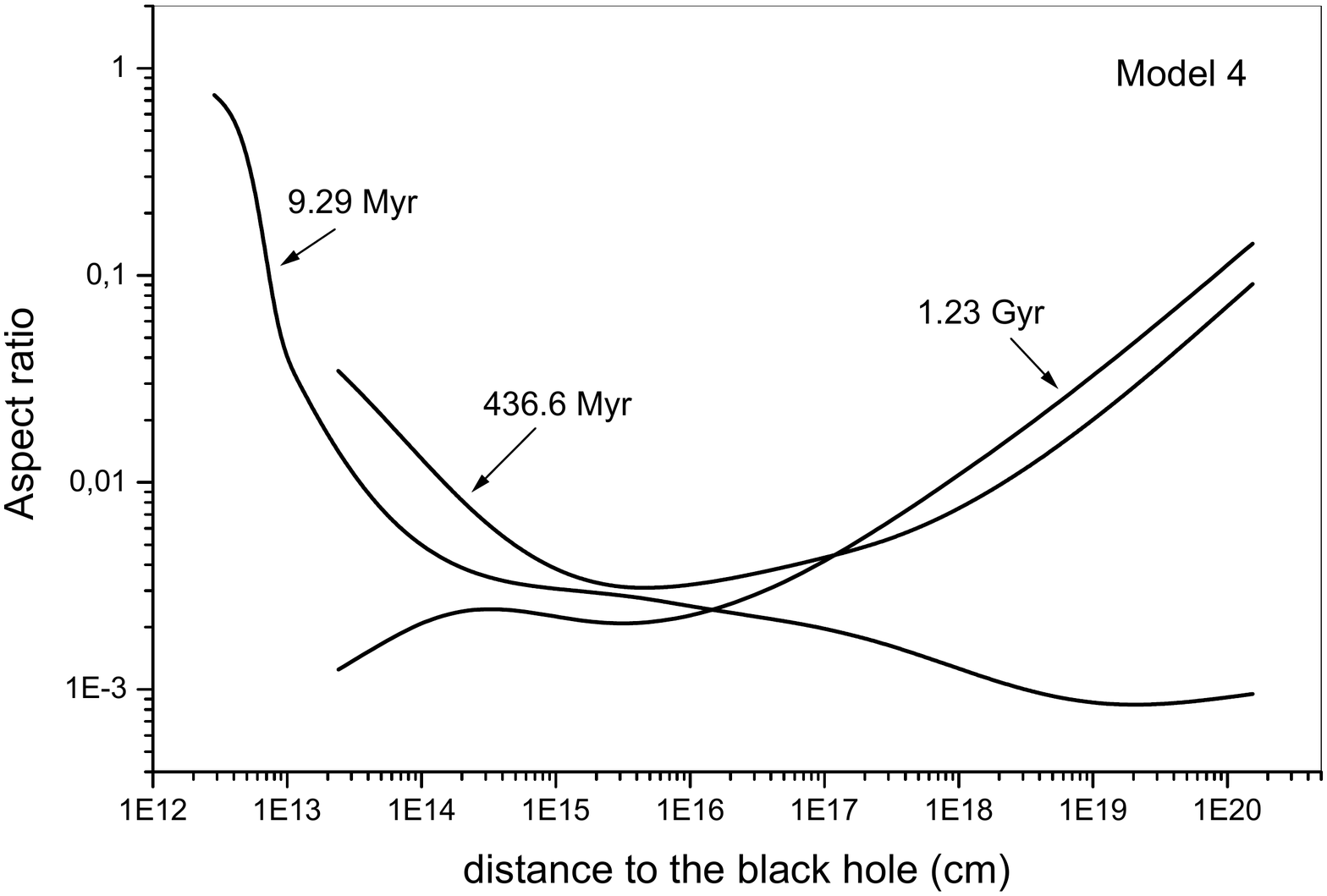}
\caption{The profile of the aspect ratio $H/r$ for model 4 in three different evolutionary stages:
9.29 Myr, 436.5 Myr and 1.235 Gyr.}
%\label{FigIV}
\end{figure}
%%%%%%%%%%%%%%%%%%%%%%%%%%%%%%%%%%%%%%%%%%%%%%%%%%%%%%%%%%%%%%%%%%%%%%%%%%%%%%%%%%%%%%%%%%%%%%%%%%

The effective optical depth controls the energy evacuation by radiation along the vertical axis and the local effective
temperature. In general, the effective optical depth increases with the distance, reflecting essentially
a decreasing gas temperature. A maximum is reached and then the optical thickness decreases quite
rapidly due to the decrease of the columnar mass density and the gas recombination. This typical behaviour
is shown in figure 5 for model 4. We can see that as the disk evolves it becomes more optically thin in
the inner region due to a decreasing columnar mass density, consequence of the accretion process. Notice 
that each curve ends at the distance at
which the disk becomes completely neutral. In this model, the ionized region increases from a 
radius of $\sim 8\times 10^{15}$ cm up to
$\sim 2\times 10^{17}$ cm in the time interval $9.3 - 436$ Myr and then decreases to $\sim 6\times 10^{16}$ cm at
the end of our calculations, i.e., $1.24$ Gyr.

%%%%%%%%%%%%%%%%%%%%%%%%%%%%%%%%%%%%%%%%%% fig 5 %%%%%%%%%%%%%%%%%%%%%%%%%%%%%%%%%%%%%%%%%%%%%%%%%%%%%%
\begin{figure}
\centering
\includegraphics[width=8cm]{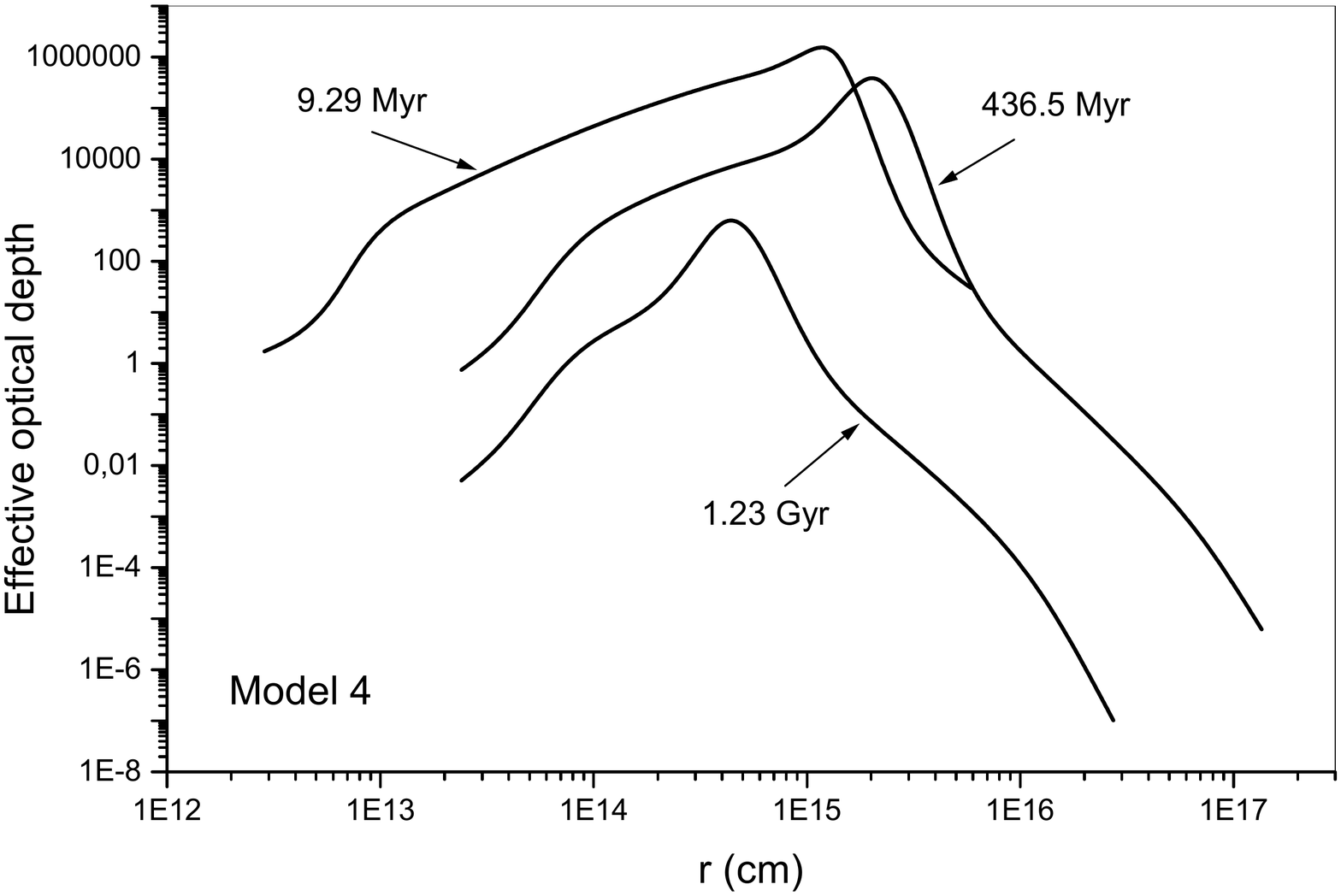}
\caption{Effective optical depth profiles for model 4 in three different evolutionary stages:
9.29 Myr, 436.5 Myr and 1.235 Gyr.}
%\label{FigV}
\end{figure}
%%%%%%%%%%%%%%%%%%%%%%%%%%%%%%%%%%%%%%%%%%%%%%%%%%%%%%%%%%%%%%%%%%%%%%%%%%%%%%%%%%%%%%%%%%%%%%%%%%

The highest effective temperatures reached in the inner region of different models after $\sim 9.3$ Myr
are of the order of $8-15$ millions K obtained respectively for models 1 and 6. These two models differ only
by the critical value of the Reynolds number and, as explained above, the former has a 
geometrical thick inner zone while the latter has a thin inner zone. Consequently, the advection
cooling is more important in model 1 than in model 6 and, in spite of having a higher viscous
dissipation, its central effective temperature is lower than that of model 6. The other models after
$\sim 9.3$ Myr have central effective temperatures of about 2 millions K, excepting model 5 which is our
``coolest" disk, since its effective temperature at that moment is only
$\sim 400,000$ K. Models 5 and 6 differ only by the initial BH seed mass and
have a geometrical thin inner scale of height as already mentioned but nevertheless their central effective
temperatures differ by a factor of four. The main reason for such a difference is found in their
mass profiles. The viscous dissipation is proportional to the columnar mass density (see eq.~\ref{viscous}) and, at
$t\sim 9.3$ Myr, $\Sigma$ in model 6 is about two orders of magnitude higher than in model 5, since the higher
seed mass of model 5 consumes faster the gas in the inner parts of the disk than in model 6. 

%%%%%%%%%%%%%%%%%%%%%%%%%%%%%%%%%%%%%%%%%%%%%%%%%% figure 6 %%%%%%%%%%%%%%%%%%%%%%%%%%%%%%%%%%%%%%%%%%%%%%%%%%%%
\begin{figure}
\centering
\includegraphics[width=10cm]{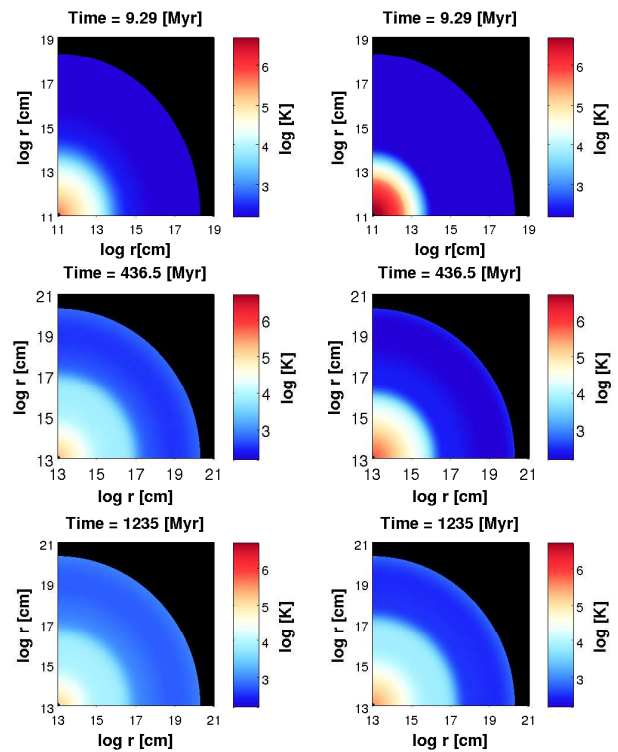}
\caption{Effective temperature maps for models 5 (left panel) and model 6 (right panel) 
in three different evolutionary stages: 9.29 Myr, 436.5 Myr and 1.235 Gyr.}
%\label{FigVI}
\end{figure}
%%%%%%%%%%%%%%%%%%%%%%%%%%%%%%%%%%%%%%%%%%%%%%%%%%%%%%%%%%%%%%%%%%%%%%%%%%%%%%%%%%%%%%%%%%%%%%%%%%%%%%%%%%%%%

Effective temperature maps for models 5 and 6 are shown in figure 6. Notice the higher
values at the inner disk reached in model 6 due to the reason already mentioned. However, the temperatures 
in the neutral zone are quite similar in both models: $T\sim 100$ K after $9.29$ Myr, $T\sim 500$ K after
$436.5$ Myr and $T\sim 670$ K after $1.235$ Gyr. The temperature in the neutral zones of disk models with
primordial chemical composition results from the balance between viscous dissipation and radiation by 
molecular hydrogen excited by collisions with H atoms. 

The evolution of the columnar mass density profile depends essentially on the inward and outward mass fluxes. The
former is controlled mainly by the critical Reynolds number and by the mass of the seed while the latter
depends mainly on the redistribution of the angular momentum throughout the disk. In other words, in the inner regions
the columnar mass density decreases as the seed grows and in the outer regions, the columnar mass density decreases
because the disk expands. Consequently, at the end of its evolution, the remaining mass of the disk is concentrated
in a ``torus-like" structure rotating around the central BH . This behaviour is shown in figure 7, in which
the evolution of the columnar mass density profile for model 3 is presented. After $\sim 1.2$ Gyr the ``torus-like"
structure having a diameter of the order of $2$ pc is clearly seen. The gas temperature in this region is about 2000 K and
it rotates with a Keplerian velocity of about 2640 km/s, controlled by the central BH mass.
It worth mentioning that in NGC 1068, the central BH is surrounded by a torus-like structure with a diameter of
$3.4$ pc, detected by the dust infrared emission (Jaffe et al. 2004). The torus-like structure is an important
element of the so-called ``unified model" of AGNs and quasars since, according to the direction of the line of sight,
it determines the visibility (or not) of the ``broad line region". It is remarkable that our model is able
to offer a natural explanation for the formation of such a ``torus-like" structure as a simple consequence of the 
disk evolution.

%%%%%%%%%%%%%%%%%%%%%%%%%%%%%%%%%%%%%%%%%%%%%%%%%%% figure 7 %%%%%%%%%%%%%%%%%%%%%%%%%%%%%%%%%%%%%%%%%%%%%%%%%%
\begin{figure}
\centering
\includegraphics[width=8cm]{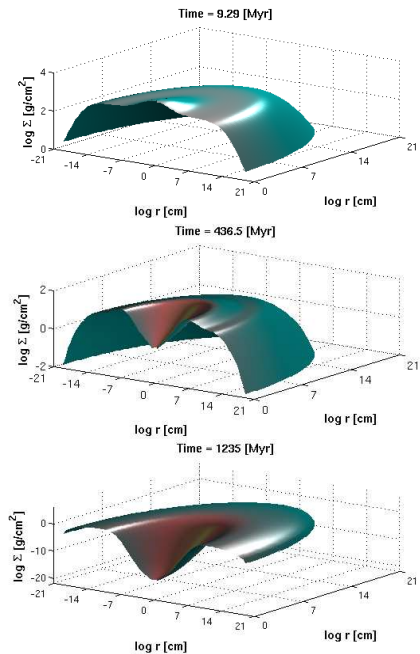}
\caption{The evolution of the columnar mass density profile: time in the different snapshots corresponds
to the same order as in previous figures. Notice the torus-like structure in the last panel, at the
end of the disk evolution.}
%\label{FigVIII}
\end{figure}
%%%%%%%%%%%%%%%%%%%%%%%%%%%%%%%%%%%%%%%%%%%%%%%%%%%%%%%%%%%%%%%%%%%%%%%%%%%%%%%%%%%%%%%%%%%%%%%%%%%%%%%%%%%%%   

\section{Conclusions}

In the present work are reported the results of numerical solutions of the hydrodynamical equations describing non
stationary accretion disks. These simulations aim to investigate the fraction of the disk mass 
transferred to the central
BH and to obtain the timescales of such a process. Turbulence generated by gravitational 
instability is supposed to be the source
of viscosity responsible for the angular momentum transport throughout the disk. The flow, self-regulated
by such a mechanism, is characterized by a critical Reynolds number with values in the range 500-1500. The 
other parameters affecting the accretion process are the initial masses of the seed and of the disk itself.

Observations of bright quasars at $z \sim 6$ associated to SMBH having masses $\simeq 10^9~M_{\odot}$ imply
that their growth by accretion processes must occur in timescales less than 900 Myr. The present investigation
suggests that our models are able to feed seeds with masses in the range $100-1500~M_{\odot}$
in the required timescale. Our numerical simulations indicate that seeds capture about a half of the original
disk mass, if the gas consumption by star formation is neglected. This conclusion is weakly dependent on the
adopted model parameters as the initial masses of the BH and of the disk or the critical Reynolds number. However,
the growth timescale, the accretion rate as well as the resulting disk luminosity do depend on the adopted 
value of those parameters. Thus, in order to obtain a quasar at $z \sim 6$ radiating with a power of 
$\sim 4\times 10^{47}$ erg/s, a disk of few $10^9~M_{\odot}$ is necessary and the accretion process should
begin around $z \sim 8$. However, it should be emphasized that in such massive disks the star formation
process is likely to be very efficient, which may reduce considerably the amount of gas available to feed
the black hole. In fact, Shlosman \& Begelman (1989) concluded from their investigation that even a low efficiency
in the gas conversion process into stars would deplete the disk on a relatively short time scale.

Although the formation of several circumnuclear disks during the history of a galaxy cannot be excluded, it is
likely that the growth of seeds is a consequence of a unique episode. In this case, one should expect that
SMBHs grow faster than their host halos. In fact, observations suggest that the growth of BHs is linked to the
evolution of galaxies but not necessarily linked to the growth of their halos. Such an ``anti-hierarchical"
growth is suggested by the fact that the co-moving number density of low X-ray luminosity AGNs peak at
$z \leq 1$ while that of high X-ray luminosity AGNs peak at $z \sim 2$ (Ueda et al. 2003; Hasinger, Miyaji \& Schmidt
2005). Since one should expect that the X-ray luminosity is directly related to the accretion rate and, consequently,
to the BH growth, these observations suggest an early assembly of the more massive objects. In fact, such 
an interpretation
is consistent with our simulations, since models with higher seed masses have higher accretion rates and higher 
luminosities. Our computations also show that the conversion efficiency of gravitational energy into radiation, measured
by the ratio $L/\dot Mc^2$, does not remain constant during the evolution of the disk, varying within the 
range $10^{-1}-10^{-3}$.

Detection of polarization in the broad emission lines of NGC 1068 provided a major argument in favour of the unified
AGN model (Antonucci 1993). According to this unified scenario, AGNs have a central core often displaying jets, a inner
broad line region (BLR), a narrow line region (NLR) and a circumnuclear torus constituted of 
dust and gas. Observation or not
of the BLR depends on the inclination angle of the line of sight with respect to the plane of the 
torus.  Recent observations (Jaffe et al. 2004) indicate that the torus is quite compact (few parsecs sized)
and having probably a ``clumpy" structure. Its origin is still under debate (see, for instance,
Elitzur \& Shlosman 2006) but such a structure appears quite naturally in our models as a consequence
of the disk evolution. The torus-like structure results from the matter consumption in the inner regions 
by the central BH and matter ``lost" in the 
outer regions as a consequence of the disk expansion by transfer of angular momentum. In the transition zone, where
the radial velocity passes from negative to positive values, the disk material remains stagnant and 
forms the aforementioned 
structure. The dimensions of the resulting torus for model 3, as mentioned above, are quite compatible 
with the parsec-sized torus observed in NGC 1068 (Jaffe et al. 2004). Moreover, the Toomre's parameter
in the torus satisfies the condition $Q \leq 1$, indicating that the gas is gravitationally unstable, which could
be an explanation for its clumpy structure, supporting our formation scenario.

Our simulations indicate that a substantial fraction of the disk remains neutral with the gas 
temperature in the interval
$100-2000$ K most of the time. This late evolution of the outer regions of accretion disk could be related, for instance,
to the molecular ``ring" of $2$ pc radius observed around Sgr $A^*$ (Gusten et al. 1987). Moreover, the 
physical conditions
prevailing in the outskirts of the disk are favourable to star formation and could be an explanation for
the presence of massive early-type stars, located in two rotating thin disks around the BH situated in our own 
galaxy (Genzel et al. 2003; Paumard et al. 2006). These stars have an estimated age of $6-8$ Myr and the total mass
under the form of stars and gas in these disks is about $2\times 10^4~M_{\odot}$. It is interesting to notice that
according to our simulations, in order to produce a SMBH of mass of about $3\times 10^6~M_{\odot}$, like the 
one located in the
Milky Way centre, a disk of initial mass of about $6\times 10^6~M_{\odot}$ is necessary. At the end of its 
evolution, it would remain
a mass of about few $10^4~M_{\odot}$, consistent with observations. This scenario will be investigated 
in more details
in a future paper reporting simulations including the possibility of gas conversion into stars.

\begin{acknowledgements}
M.A.M.A. acknowledges the Comisi\'on Nacional de Investigaci\'on Cient\'ifica y
Tecnol\'ogica (CONICYT) de Chile for the fellowship which permitted his
stay at the Observatoire de la C\^ote d'Azur.     
\end{acknowledgements}

\end{document}